\documentclass[reprint,superscriptaddress, amsmath,amssymb, aps, pra, longbibliography]{revtex4-1}

\usepackage{soul}
\usepackage{amsmath}
\usepackage{amssymb}
\usepackage{graphicx}
\usepackage{dcolumn}

\usepackage{bm}
\usepackage{amsmath}
\usepackage{graphicx}
\usepackage{amsfonts}
\usepackage{outlines}
\usepackage{graphicx}
\usepackage{array}
\usepackage{float}
\usepackage{color}
\usepackage{amssymb}%
\usepackage[subnum]{cases}
\usepackage[colorlinks=true,linkcolor=blue]{hyperref}%
\hypersetup{allcolors=blue}
\usepackage[normalem]{ulem}
\usepackage{xcolor}
\renewcommand{\Re}{\mathrm{Re}}

\newcommand{\qty}[2]{\ensuremath{#1\,\text{#2}}} 

\newcommand{\opt}{\mathrm{opt}}
\newcommand{\tgt}{\mathrm{tgt}}
\newcommand{\Op}[1]{\ensuremath{\mathsf{\hat{#1}}}}

\newcommand{\ket}[1]{\ensuremath{\left\vert #1 \right\rangle}}
\usepackage{mathtools}
\newcommand{\Avg}[1]{\ensuremath{\left\langle #1 \right\rangle}}
\DeclarePairedDelimiterX\braket[2]{\langle}{\rangle}{#1 \delimsize\vert #2}

\newcommand{\micro}{\textmu{}}
\newcommand{\micrometer}{\micro{m}}
\newcommand{\microsecond}{\micro{s}}
\newcommand{\pps}{$\pi$/s}

\newcommand{\dx}{\Delta x}
\newcommand{\tLoop}{t_{\text{loop}}}
\newcommand{\ie}{{\sl{\mbox{i.e.}}}}
\newcommand{\eg}{{\sl{\mbox{e.g.}}}}

\begin{document}

\title{Rotation sensing using tractor atom interferometry}
\date{\today }

\author{Bineet Dash}
\email{bkdash@mich.edu}
\affiliation{Department of Physics, University of Michigan, Ann Arbor, MI 48109}
\author{Michael H. Goerz}
\affiliation{DEVCOM Army Research Laboratory, 2800 Powder Mill Road, Adelphi, MD 20783}
\author{Alisher Duspayev}
\affiliation{Department of Physics, University of Michigan, Ann Arbor, MI 48109}
\author{Sebasti\'an C. Carrasco}
\affiliation{DEVCOM Army Research Laboratory, 2800 Powder Mill Road, Adelphi, MD 20783}
\author{Vladimir S. Malinovsky}
\affiliation{DEVCOM Army Research Laboratory, 2800 Powder Mill Road, Adelphi, MD 20783}
\author{Georg Raithel}
\affiliation{Department of Physics, University of Michigan, Ann Arbor, MI 48109}

\begin{abstract}
We investigate a possible realization of an ultracold-atom rotation sensor that is based on recently proposed tractor atom interferometry (TAI). An experimental design that includes generation of a Laguerre-Gaussian-beam-based ``pinwheel'' optical lattice and multi-loop interferometric cycles is discussed. Numerical simulations of the proposed system demonstrate TAI rotation sensitivity comparable to that of contemporary matter-wave interferometers. We analyze a regime of TAI rotation sensors in which nonadiabatic effects may hinder the system's performance. We apply quantum optimal control to devise a methodology suitable to address this nonadiabaticity. Our studies are of interest for current efforts to realize compact and robust matter-wave rotation sensors, as well as in fundamental-physics applications of TAI.
\end{abstract}

\maketitle

\section{Introduction}%
\label{sec:intro}

Recent progress in atom interferometry (AI) has raised promising prospects in fundamental physics~\cite{Tarallo2014, Schlippert2014, Kovachy2015, Jaffe2017, Rosi2017}, precision measurements~\cite{Fixler74, Parker191, Xu745, morel2020} and practical applications~\cite{Bongs2019} including geodesy, seismology and inertial sensing with atomic acceleration and rotation sensors. Focusing on rotation, the interferometric measurement relies on the Sagnac phase $\phi_s = 2 E A/hc^2$ arising between wave-packets of energy $E$ that are counter-rotating around an area $A$. Since their first demonstration in 1913~\cite{Culshaw_2006}, optical Sagnac interferometers have achieved sensitivities beyond $10^{-10}$\,rad/s in fiber-optic gyroscopes (FOGs) and large-area pinwheel laser-based setups. The motivation to design Sagnac atom interferometers stems from the potential orders-of-magnitude enhancement in sensitivity that scales inversely with the associated de Broglie wavelength~\cite{croninreview}.

Previous experiments and proposals for the realization of Sagnac AIs include free-space~\cite{riehle1991, gustavson1997, Schubert2021, BARRETT2014875} and point-source interferometers~\cite{dickerson, hoth, chen2020}, where atomic fountains or dropped atomic clouds propagate freely along interfering paths, as well as guided-wave AIs~\cite{wu2007demonstration, moan2020, beydler2023,krzyzanowska2022matter}. Despite their much smaller particle flux and interferometric areas, these designs have recently surpassed the sensitivity of FOGs. However, free-space AIs can be space- and power-intensive, as their sensitivity scales as the interrogation time squared~\cite{croninreview}, fueling a push to increasing drop heights and apparatus sizes in earth-based experiments. In order to achieve higher sensitivity combined with compact setups, multi-pass guided-wave designs have been proposed based on trapped ions~\cite{campbell2017rotation}, weak magnetic traps~\cite{wu2007demonstration, qi2017magnetically,muruganandam2003bose}, time-averaged adiabatic potentials~\cite{stevenson2015sagnac, gentile2019ring, Ryu_2015, bell2016bose}, toroidal optical traps~\cite{wang2009atom} and optical waveguide formed by collimated laser beams~\cite{krzyzanowska2022matter}.

The performance of free-space and atom-guide AIs is often limited by the dispersion of the atomic wave functions along unconfined degrees of freedom, inefficient closure of interferometric paths,
and Landau-Zener tunneling in spinor implementations
~\cite{stickney2002, stickney, jo2007}. Tractor atom interferometry (TAI)~\cite{duspayev_tractor_2021}, a recently proposed technique, seeks to address these issues by uninterrupted three-dimensional confinement and transport of atomic wave packets along programmable trajectories using optical or other traps. Robust AI implementations for acceleration sensing using deep, spin-dependent optical potentials and optical tweezers have been explored in recent proposals~\cite{steffen_digital_2012,duspayev_tractor_2021, raithel2022principles, premawardhana2023}.

In this paper, we present investigations on a possible realization of a rotation sensor using TAI. Our azimuthal optical lattice and its matter-wave Hamiltonian are outlined in Sec.~\ref{sec:design}. Aspects of the interferometer operation and its matter-wave dynamics are explained in Sec.~\ref{sec:operation}. From our numerical quantum-dynamics simulations presented in Sec.~\ref{sec:numerics} we infer the sensitivity and confirm agreement with semiclassical predictions that apply in the adiabatic limit. In Sec.~\ref{sec:nonadiabatic} we then quantify and discuss possible nonadiabatic excitations during operation of the TAI interferometer. Finally, results that incorporate  the application of optimal control theory to minimize detrimental nonadiabatic effects are presented in Sec.~\ref{sec:optimalcontrol}. The paper is concluded in Sec.~\ref{sec:conclusion}.

\section{Pinwheel optical lattice design}
\label{sec:design}

\begin{figure}[tb]
  \includegraphics[width=0.4\textwidth]{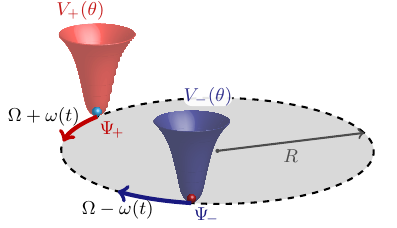}
  \caption{%
    Concept of TAI as a Sagnac rotation sensor. Split  azimuthal atomic trapping potentials $V_{+}(\theta)$ and $V_{-}(\theta)$ containing coherently-split wave-packet components $\Psi_{\pm}(\theta)$, represented here as well-localized dots, are counter-rotated in the instrument's rest frame at tunable angular velocities $\pm \omega(t)$ along counter-wound circular trajectories of radius $R$. This occurs in the presence of a background angular velocity $\Omega$ of the instrument's rest frame against an inertial frame, in which the rotation speeds are $\Omega \pm \omega(t)$. The value of $\Omega$ is to be measured.
    }
    \label{fig:concept-fig}
\end{figure}

As depicted in Fig.~\ref{fig:concept-fig}, the principles of spinor-TAI~\cite{duspayev_tractor_2021} can be leveraged for rotation sensing by designing circular trajectories along which spin-dependent potentials carry trapped atomic wave-function components in opposite directions. The potentials must be designed to strongly confine the trapped wave functions in all spatial dimensions to minimize nonadiabatic effects and dispersion. The trajectory pairs are closed and cover a half- or full-integer number of loops in each trajectory.
This can be realized by a pair of deep, spin-dependent, counter-rotating ``pinwheel" optical lattices.
Such lattices can be created using co-propagating Laguerre-Gaussian (LG) beams~\cite{PhysRevLett.78.4713, franke-arnold_optical_2007}, interference of Gaussian and hollow beams with a quadrupole magnetic moment~\cite{courtade_dark_2006}, or interference of LG beams with plane waves in the presence of a conical magnetic field~\cite{amico_quantum_2005} for twisted boundary conditions. Here we focus on the first approach, which is an all-optical technique suitable to create both bright (red-detuned) and dark (blue-detuned) lattices with several, widely tunable parameters.

The electric field of an LG beam with azimuthal index $l$, zero radial index, frequency $f_l$ and wave vector $k_l$ propagating along the positive $z$-direction is given in phasor notation and cylindrical coordinates $\mathbf{r}\equiv \left(r, \theta, z \right) $ as
\begin{equation}
    LG_l(\mathbf{r}, t)
      =
        \frac{\mathcal{E}_l(r,z)}{\sqrt{c \epsilon_0}}\,
        e^{i \left[2\pi f_l t +\Phi_l(z) + l\theta -k_l\left(z+\frac{r^2}{2R(z)}\right)\right]}\,,
\end{equation}
 where $c$ and $\epsilon_0$ are the speed of light and vacuum permittivity, respectively, and with the amplitude
\begin{equation}
  \mathcal{E}_l(r,z)
  = 
   \sqrt{\frac{4P}{\pi \vert l \vert! w(z)^2 }}
    \left(\frac{\sqrt{2} r}{w(z)}\right)^{|l|} e^{-\frac{r^2}{w(z)^2}}\,.
\end{equation}
$P$ is the laser beam power and $w(z) = w_0 \sqrt{1+(z/z_R)^2}$
is the beam-waist parameter with the Rayleigh range $z_R = \pi w_0^2/\lambda$. The radius of the phase front's curvature is $R(z) = z\left(1+(z_R/z)^2\right)$, and $\Phi_l(z) = (|l|+1) \arctan\left(z/z_R\right)$ is the Gouy phase. Along the $z$-axis, the LG beam has an optical vortex line featuring a phase singularity and vanishing intensity. Due to the azimuthal ($\theta$) phase dependence, the interference of two co-propagating LG beams with modes $l_1$ and $l_2=l_1+m$ and frequency $f_1$ and $f_2=f_1-\Delta f$ results in an intensity distribution
\begin{widetext}
\begin{equation}
\begin{split}
  \vert\mathcal{E}\vert^2(\mathbf{r}, t)
   &=    \mathcal{E}^2_{l_1}(r,z) + \mathcal{E}^2_{l_2}(r,z) \\
   & \qquad
   + 2 \mathcal{E}_{l_1}(r,z)\mathcal{E}_{l_2}(r,z)\cos\left[2\pi(\Delta f) t - \Delta \Phi(z) - \frac{2\pi (\Delta f) z}{c} - \frac{\pi r^2}{c}\left(\frac{f_1}{R_1(z)}- \frac{f_2}{R_2(z)}\right) - m \theta \right]\,.
\label{eq:approxLGpotential}
\end{split}
\end{equation}
\end{widetext}

The salient feature of this interference pattern is the sinusoidal modulation of intensity in the azimuthal ($\theta$) direction, in the very last term. In experimentally relevant cases, \eg, for a pinwheel optical lattice of radius \qty{10-100}{\micrometer} rotating at \qty{10-1000}{Hz}, the terms proportional to $\Delta f/c$ and $r^2/c$ inside the cosine are negligible. The difference in Gouy phase, $\Delta \Phi(z) = m\arctan(z/z_R)$, in principle twists the pinwheel azimuthally as a function of $z$. However, as described in the following, the atoms are further trapped along the $z$ direction by a separate, far off-resonant one-dimensional static optical lattice with lattice planes extending transverse to $z$. The twisting angle due to the variation of $\Delta \Phi(z)$ within one spatial period of the static $z$-lattice is typically less than \qty{1}{mrad} and is therefore negligible. With these approximations, the optical potential near $z=0$ reads
\begin{equation}
  V(\mathbf{r}, t)
    \approx   V_1(r) + V_2(r)\cos\left[2\pi(\Delta f) t - m \theta \right]\,,
\label{eq:lgintensity}
\end{equation}
with $V_1(r) = -\frac{\alpha}{2c\epsilon_0}\left(\mathcal{E}^2_{l_1}(r,0) + \mathcal{E}^2_{l_2}(r,0) \right)  $ and $V_2(r) = -\frac{\alpha}{c\epsilon_0} \mathcal{E}_{l_1}(r,0)\mathcal{E}_{l_2}(r,0)$, where $\alpha$ is the polarizability of the selected atomic state. Therefore, a pair of LG beams with a small detuning of $\Delta f$ and with $l$-indices differing by $m$ effectively create a pinwheel optical lattice with $m$ azimuthal lattice sites, rotating at a tunable angular velocity $\omega= 2\pi\Delta f/m$.

A counter-rotating pinwheel lattice of similar size can be obtained from a second pair of LG beams with opposite detuning that are superimposed over the first pair. The pinwheel lattices can be made spin-selective by tuning the wavelengths of the beam pairs forming the lattice so that they trap different atomic spin states. A particular example of such spin states are the $\ket{5S_{1/2}, F=1, m_F=0}$ and $\ket{5S_{1/2}, F=2, m_F=0}$ states of $^{87}$Rb, with the wavelengths of the respective pinwheel lattices set near the $D_1$ line ($\approx$~\qty{795}{nm}), as described in Ref.~\cite{raithel2022principles}.

Trapping in such lattices requires blue-detuned light, where the
atoms are trapped near intensity minima. In that case, the photon scattering rate of the atoms in the lattice light is minimal, thereby minimizing both photon-scattering-induced decoherence of the interferometer as well as decoherence caused by trap-laser intensity fluctuations. The beam parameters for $LG_{l_1}$ and $LG_{l_2}$-modes in Eq.~(\ref{eq:lgintensity}) must be chosen carefully to create a sufficiently deep and tightly confined radial potential in order to suppress wave-function dynamics in the radial direction. At the same time, this potential must go through a zero-intensity minimum to trap atoms with minimal coherence loss due to photon scattering. This can be achieved when both LG beams have similar maximum intensity, with the radial intensity maxima separated by more than one FWHM of the radial intensity distributions. For a Gaussian beam waist ratio $\eta = w_{0,2}/w_{0,1}$ and a power ratio
$P_{2}/P_{1} = \eta^2 \sqrt{l_2/l_1}$, the radial intensity maxima are similar and separated by $\Delta r \approx w_{0,1}\left(\eta \sqrt{l_1} - \sqrt{l_2}\right) / \sqrt{2}$ near the focus. Since the FWHM of $LG_{l_i}$ is on the order of $w_{0,i}$, for a given $l_1$ and $l_2$, the waist ratio $\eta$ should be chosen such that  $\Delta r \gtrsim (1+\eta) w_{0,1}$. The number of desired azimuthal wells in the pinwheel lattice, $m$, has an implicit effect
on the best choice for $\eta$ because $m= \vert l_2-l_1 \vert$. We have found that $\eta \sim 1.3-1.8$ works for most $l_1$ and $l_2$ $< 50$.

\begin{figure}[tb]
 \centering
  \includegraphics[width=0.48\textwidth]{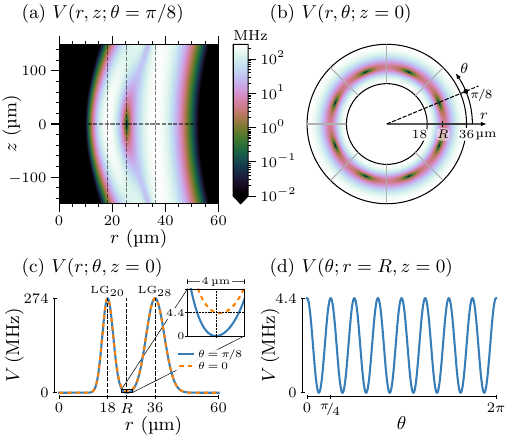}
  \caption{
    \label{fig:pinwheel-lattice-potential}
    An 8-site pinwheel $^{87}$Rb optical lattice is created from LG$_{20}$ and LG$_{28}$ laser modes with \qty{\lambda=795}{nm} and
    respective Gaussian waists of $w_{0,1}=$ \qty{5.78}{\micrometer} and $w_{0,2}=$ \qty{9.66}{\micrometer},
    and powers of $P_1=$ \qty{5.38}{mW} and $P_2=$ \qty{17.78}{mW}. The beams are assumed to be blue-detuned for an AC polarizability of \qty{-0.16}{Hz/(V/m)$^2$}.
    (a) Optical potential in the $r-z$ plane.
    (b) Optical potential in the transverse plane at the focus $z=0$.
    (c) Potential along the radial direction for $z=0$ and $\theta = 0$ and $\pi/8$.
    (d) Potential along the azimuthal direction at \qty{R = 25.46}{\micrometer} and $z = 0$.
  }
\end{figure}
For the example of $^{87}$Rb, Fig.~\ref{fig:pinwheel-lattice-potential}(a) demonstrates the superposition of two \qty{795}{nm} laser beams with modes $LG_{20}$ and $LG_{28}$. In this case, $\eta=1.67$ leads to an ideal pinwheel optical lattice with 8 sites, as shown in the optical potential in the transverse plane in Fig.~\ref{fig:pinwheel-lattice-potential}(b). The trapping potential is about \qty{4.4}{MHz} deep and perfectly sinusoidal along the azimuthal direction, acting as a lattice with periodic boundary conditions. Ultracold $^{87}$Rb atoms can be trapped in these potential minima with minimal photon scattering~\cite{raithel2022principles}. As described below, this potential is sufficiently deep to prevent wave-function dispersion or tunneling between the lattice sites. Along the radial direction, the potential is about \qty{274}{MHz} deep, and the radial trap frequency approximately equals 5 times the azimuthal trap frequency, $\omega_r \approx 5 \omega_{\theta}$.

Next, we discuss TAI confinement in the axial ($z$) direction. Other experiments~\cite{amico_quantum_2005, courtade_dark_2006} on ring-like traps have reported axial confinement using lattices created by counter-propagating laser modes. In the case of rotating pinwheel lattice, superposition of detuned counter-propagating LG beams can lead to unwanted axial movement. Therefore, we here suggest co-propagating LG beam pairs to form the pinwheel lattice, and to use a separate, far off-resonant one-dimensional optical lattice along the $z$-direction using counter-propagating Gaussian beams of a sufficiently large beam waist. This allows robust, all-optical axial confinement of the atoms on the pinwheel. For example, a \qty{300}{kHz}-deep optical lattice can be created by counter-propagating $\approx$\qty{1}{W} \qty{532}{nm}-wavelength Gaussian beams, focused to a waist of \qty{100}{\micrometer}. This will generate an axial stack of many pinwheel lattices spaced by an axial lattice period of \qty{266}{nm}. The structure of the pinwheel lattices, as shown in Fig.~\ref{fig:pinwheel-lattice-potential}(b)-(d), remains largely constant over an axial range of about \qty{5}{\micrometer} from the focus, suggesting that several tens of near-identical pinwheel lattices with tight 3D confinement can be stacked.

With the radial and axial degrees of freedom being essentially frozen, the pinwheel optical lattices can be approximated as 1D lattices with periodic boundary conditions. Assuming that there is no linear background acceleration, the Hamiltonian of the system in a suitable inertial frame can be written as
\begin{equation}
  H_{\pm}(\theta, t) = -\frac{\hbar^2}{2I}\frac{\partial^2}{\partial \theta^2} + V_0 \cos\left(m \theta + \phi_{\pm}(t) \right)\,.
  \label{eq:hamiltonian}
\end{equation}

This expression is in the coordinate representation, and the labels ``$+$'' and ``$-$'' refer to the two atomic spin states (which are rotated in opposite directions). The kinetic term contains the effective moment of inertia $I=m_{\text{Rb}} R^2$ of a $^{87}$Rb atom (atomic mass $m_{Rb}$) rotating on a pinwheel of radius $R$. The radius $R$ is defined as the center of mass of the tightly-confined radial wave function. The cosine potential with $m$ sites moves with phases
\begin{equation}
  \phi_{\pm}(t)
  = \int_{0}^{t} \omega_{\pm}(t^\prime)\,dt^\prime
  = \int_{0}^{t}\left( \Omega \pm \omega(t^\prime)\right)\,dt^\prime\,.
  \label{eq:phase}
\end{equation}
The phases are controlled via the tunable angular velocity $\omega(t)$. The goal of the present TAI scheme is to measure the constant rotation rate $\Omega$ of the instrument's rest frame (``lab frame'') against the inertial frame.

The Hamiltonian in Eq.~(\ref{eq:hamiltonian}) includes the effect of the Euler force, while other non-inertial forces like centrifugal and Coriolis force can be neglected in this scheme. During interferometer operation, the lattices are rotated much slower ($\omega_{\pm} \lesssim \qty{10^2}{\pps} $) than the radial trap frequency ($\omega_r \sim \qty{10^4}{\pps}$). In this regime, the relative displacement due to the centrifugal force, $\Delta R/R \simeq (\omega_\pm/\omega_r)^2$, is negligible, and the lattice radius $R$ can be assumed to be constant throughout. As shown in Sec.~\ref{sec:numerics},
in the desired adiabatic regime the wave packets are at rest in frames that co-rotate with the pinwheel lattices, and therefore do not experience any Coriolis effect.

\section{Operation}%
\label{sec:operation}

The interferometer is initialized by co-aligning the axes and azimuthal minima of the spin-dependent pinwheel lattices for the pair of utilized spin states. The internal spin states could correspond, \eg, to the $\ket{+} = \ket{5S_{1/2},F=1,m_F=0}$ and $\ket{-} = \ket{5S_{1/2},F=2, m_F=0}$ states in the ground-state manifold of $^{87}$Rb.
At $t=0$, the wave function is prepared in the local ground state $\Psi_0(\theta)$ of one particular site in the ``$+$''-lattice, that is, $\ket{\Psi(\theta, t=0)} = \Psi_0(\theta) \ket{+}$.
The shape and the width of the eigenstate depend on the radius implicit in the effective moment of inertia $I$, the number of lattice sites, $m$, and the $V_0$ of the trapping potential.
For a given $m$ and a sufficiently deep potential (large $V_0$), the wave packet $\Psi_0(\theta)$ will be close to the eigenstate of a quantum harmonic oscillator with frequency
\begin{equation}
  \label{eq:omega_H}
  \omega_\theta = m\sqrt{ V_0 / I}\,,
\end{equation}
obtained by a Taylor expansion of the potential in Eq.~\eqref{eq:hamiltonian} at the first site. In the remainder of the paper, we will consider optical lattices with a radius of $R=\qty{25.46}{\micrometer}$ and $m=8$ lattice sites, which corresponds to a lattice period of \qty{10}{\micrometer} in the azimuthal direction. Without loss of generality, we choose lattice phases such that the initial wave packet is centered at $\theta_0 = \pi/8$ at the first site.

Driving a momentum-transfer-free optical Raman transition at a suitable Rabi frequency $\Omega_{\pm}$, we  implement a $\pi/2 $-pulse,
\begin{equation}
  \label{eq:U_pihalf}
  \Op{U}_{\pi/2} = \frac{1}{\sqrt{2}}\begin{pmatrix}
    1 & i \\
    i & 1
  \end{pmatrix}\,,
\end{equation}
between the two spin components. This acts as a beam-splitter and creates an equal superposition of the two spin states. Thus, the wave packets in the two spin-dependent potentials immediately after the $\pi/2$-pulse are
\begin{equation}
  \Psi_{+}(\theta, 0) = \frac{1}{\sqrt{2}} \Psi_0(\theta)\,,
  \quad
  \Psi_{-}(\theta, 0) = \frac{i}{\sqrt{2}} \Psi_0(\theta)\,.
\end{equation}
The duration of the $\pi/2$-pulse typically is negligible compared to the overall duration of the interferometer sequence. An experimentally suitable choice
for the duration of the  $\pi/2$-pulse could be, for instance, \qty{1.4}{\microsecond}, corresponding to a Rabi frequency of $\Omega_{\pm}= \pm \qty{2\pi\times 178}{kHz}$.

After splitting, the two wave packets $\Psi_{\pm}(\theta, t)$ evolve independently (\ie, without spin coupling) under the Hamiltonian in Eq.~\eqref{eq:hamiltonian} with counter-rotating time-dependent angular velocities $\pm \omega(t)$.
For the time being, we assume that $\omega(t)$ varies sufficiently slowly for the wave-packet evolution to be adiabatic, \ie,  the $\Psi_{\pm}(\theta, t)$ remain in the ground state of the local lattice site potential at all times.
For the azimuthal ramp of the pinwheel lattices, here we first choose the smoothly varying function

\begin{numcases}{%
  \omega(t) = \label{eq:drive}
  }
  \omega_{0} \sin^2\left(\frac{\pi t}{2 t_r}\right)         & $0 \leq t < t_r$                                           \label{eq:drive_up} \\
  \omega_{0}                                                & $t_r \leq t < t_r + \tLoop$                       \label{eq:drive_free}\\
  \omega_{0} \cos^2\left(\frac{\pi t^\prime }{2t_r} \right) & $T - t_r \leq t \leq T$ \label{eq:drive_down}\,,
\end{numcases}
with $t^\prime = t-t_r-\tLoop$ and the total duration $T=2t_r + \tLoop$.

During the ramp-up time $t_r$, the angular speeds of the lattice potentials in the instrument frame are accelerated from $\pm \omega(0) = 0$ to $\pm \omega(t_r) = \pm \omega_0$, and subsequently remain constant for a duration of $t_{loop}$. We first assume that $t_r$ is sufficiently large to result in adiabatic dynamics. After the loop time $t_{loop}$, the lattices are decelerated from $\pm \omega_0$ to $\omega(T) = 0$ by running the ramp-up control backwards. At final time $T$, the two lattice potentials and, thus, the final wave packets must coincide (both in the instrument frame (lab frame) and in the lattice rest frames). This is achieved by adjusting $\tLoop$ such that
\begin{equation}
  \label{eq:loop_condition}
 \int_{0}^{T}\omega(t)\,dt = n \, \pi
\end{equation}
for an interferometer with $n$ ``cycles''. The effective area of the TAI then equals $n \times \pi R^2$.

\begin{figure}[tb]
 \centering
  \includegraphics{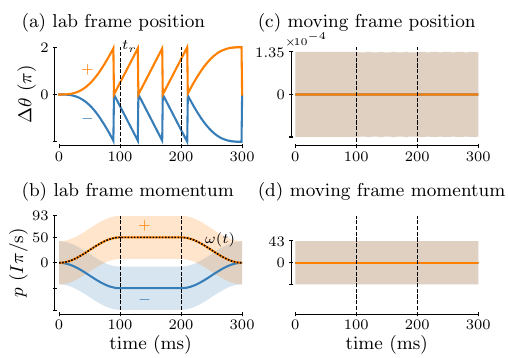}
  \caption{%
    \label{fig:adiabatic_dynamics}
    Adiabatic wave-function evolution in a rotating pinwheel optical lattice using $\omega(t)$ from Eq.~\eqref{eq:drive} with $V_0 = \qty{h\times 4.4}{MHz}$, $t_r = \tLoop = \qty{100}{ms}$, $\omega_0 = \qty{50}{\pps}$, and $\Omega=0$.
    (a) Expectation value of the azimuthal displacement of the wave packets relative to the initial position $\theta_0 = \pi/8$ of the ground state of the selected pinwheel-lattice well, $\Delta\theta \equiv \langle \theta \rangle - \theta_0$, as measured in the instrument's rest frame (``lab frame'') for both counter-rotating states $\Psi_{+}(t)$ and $\Psi_{-}(t)$. The lattice has \qty{R = 25.46}{\micrometer} and $m = 8$~lattice sites.
    (b) Momentum expectation value in the lab frame, in units of effective moment of inertia $I=m_{\text{Rb}} R^2$ of a $^{87}$Rb atom rotating
    at \qty{1}{\pps}. The control function for the angular velocity of the pinwheel-lattices, $\omega(t)$, is shown as the dotted black curve.
    (c) Azimuthal-angle expectation value of the wave packet components in the lattices' moving frames,\ie, the frames in which the lattices are stationary.
    (d) Angular-momentum expectation value in the moving frame, \ie, average angular momentum relative to $\pm I \omega(t)$.
    The shaded regions in panels (b--d) indicate the standard deviations of the respective wave-function densities, that is, the widths of the wave packets.
  }
\end{figure}

Some exemplary dynamics for adiabatic evolution under Eq.~\eqref{eq:drive} are shown in Fig.~\ref{fig:adiabatic_dynamics}. The interferometer has $n=10$ cycles, as can be seen in panel~(a). The expectation value of the momentum, seen in panel~(b), follows exactly the movement of the potential, controlled by $\omega(t)$. In the moving frames, defined here as the rest frames of the rotating pinwheel lattices, the wave packets remain perfectly stationary, see panels~(c, d).

The interferometric scheme is completed at final time $T$ by an inverse $\pi/2$-pulse, denoted $\Op{U}_{\pi/2}^\dagger$, see Eq.~\eqref{eq:U_pihalf}, to recombine the two spin-dependent components. For a non-zero constant background rotation $\Omega$ in Eq.~\eqref{eq:phase}, the wave packets $\Psi_{+}(\theta, T)$ and $\Psi_{-}(\theta, T)$ accumulate a differential phase $\Delta \Phi$ that is reflected in the recombined state
\begin{equation}
  \Op{U}_{\pi/2}^\dagger \ket{\Psi(T)} = c_{+}(\theta, T) \ket{+} + c_{-}(\theta, T) \ket{-}
\end{equation}
with the populations
\begin{equation}
  \label{eq:pop_nonadiabatic}
  |c_{\pm}|^2
  = \frac{1}{2}
    \pm \frac{1}{2} \Re\left[\eta e^{-i \Delta\Phi}\right]
\end{equation}
and the overlap of the final-time wave-packet components
\begin{equation}
  \label{eq:eta_overlap}
  \eta = \braket{\Psi_{-}(\theta, T)}{\Psi_{+}(\theta, T)} \,.
\end{equation}

For a closed interferometric path and adiabatic time evolution, $\Psi_{-}(\theta, T) = \Psi_{+}(\theta, T) = \Psi_0(\theta)$, and thus $\eta = 1$. In this case, Eq.~\eqref{eq:pop_nonadiabatic} simplifies to
\begin{subequations}%
  \label{eq:pop_adiabatic}
  \begin{align}
    |c_{-}|^2 &= \frac{1}{2} - \frac{\cos{\Delta\Phi}}{2} = \sin^2\left(\frac{\Delta\Phi}{2}\right) \,, \label{eq:pop_minus}\\
    |c_{+}|^2 &= 1 - |c_{-}|^2 = \cos^2\left(\frac{\Delta\Phi}{2}\right)\,.\label{eq:pop_plus}
  \end{align}
\end{subequations}
Up to an offset of an integer multiple of $\pi$, the value of $\Delta \Phi$ can be derived from a measurement of the population in at least one of the two spin states.

\section{Numerical Simulation}%
\label{sec:numerics}

\subsection{Quantum methods}%
\label{subsec:quantum}

The Crank-Nicolson (CN) method~\cite{muruganandam2003bose,PhysRevA.79.043630,PhysRevA.102.053312} has commonly been employed for simulations of wave packet dynamics in the position representation, including cases with moving potentials~\cite{duspayev_tractor_2021}. The method requires a computationally expensive ($\mathcal{O}(N_x^3)$) matrix inversion. In practice, for systems with 1D scalar potentials and non-PBC, this is usually reduced to $\mathcal{O}(N_x)$ due to the tridiagonal structure of the Hamiltonian in position space. In the context of the azimuthal optical lattice, periodic boundary conditions introduce additional corner entries in the Hamiltonian matrix, necessitating a generalized Crout reduction, as explained in Appendix~\ref{sec:pbc-Crout}.

\begin{figure}[tb]
 \centering
  \includegraphics[width=0.5\textwidth]{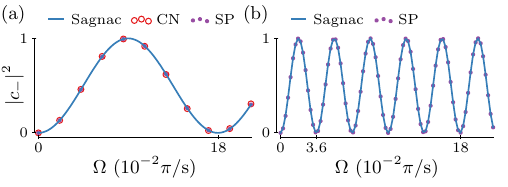}
  \caption{%
    \label{fig:cn-sim-results}
    Interferometric response of TAI in a pinwheel optical lattice to a constant background rotation $\Omega$ for $V_0 = h\times \qty{2.2}{MHz}$, $t_r=\qty{100}{ms}$, and $T = \qty{300}{ms}$ (other parameters see text).
    (a) Population in $\ket{-}$ for a recombination after $n=2$ cycles,  with $\omega_0 = \qty{10}{\pps}$. The Sagnac curve is analytically calculated from Eq.~\eqref{eq:pop_minus} with $\Delta\Phi = \Delta\Phi_S$, Eq.~\eqref{eq:sagnac}. The ``CN'' and ``SP'' points are obtained from simulations of the quantum dynamics with the Crank-Nicolson and split-propagator methods, respectively (see text for details).  (b) Population in $\ket{-}$ after n=10 cycles with $\omega_0 = \qty{50}{\pps}$, cf.\ the dynamics in Fig.~\ref{fig:adiabatic_dynamics}.
  }
\end{figure}

Here, we use CN simulations to study a TAI in a pinwheel optical lattice as a function of $\Omega$. The results of CN simulations performed in the inertial frame according to the Hamiltonian in Eq.~(3) are shown in Fig.~\ref{fig:cn-sim-results}~(a) as the points labeled ``CN", with the parameters listed in the figure caption. A time-step $\Delta t= \qty{50}{ns}$ and 3200 spatial grid points for the full range of $\theta \in [0, 2\pi]$ have been used. From the simulations, we verify that for parameters as in Fig.~\ref{fig:cn-sim-results}, a $h \times$\qty{4.4}{MHz} deep pinwheel lattice effectively prevents any tunneling between the lattice sites. Consequently, the wave-packet dynamics in the co-rotating frames of reference is confined within one lattice site, or equivalently within a $\theta$-range of only $2 \pi/m$ in width. Exploiting the localization of the wave-packet components in their respective lattices, in the present case the spatial grid can be reduced in width by a factor of $m=8$ to the region $[0, \pi/4]$ of a single lattice site by applying the transformation

\begin{equation}
  \begin{split}
    \hat{U}_{\pm}(t)
    &=
      \exp\left(\frac{-i\hat{L}_z \phi_{\pm}(t)}{\hbar}\right) \\
    &\equiv
      \exp\left(-\int_0^{t}\!\omega_{\pm}(t^\prime)\,dt^\prime\,\frac{\partial}{\partial \theta}\right)
  \end{split}
  \label{eq:transform}
\end{equation}
\noindent into the lattices' rest frames, in which $\theta$ is relative to the moving lattice potentials. Applying the transformation in Eq.~\ref{eq:transform} on the Hamiltonian in Eq.~\eqref{eq:hamiltonian}, one finds the Hamiltonian in the lattice rest frames,
\begin{equation}
    \tilde{H}_{\pm} (t)= -\frac{\hbar^2}{2I}\frac{\partial^2}{\partial \theta^2} + V_0 \cos\left(m \theta\right) - i \hbar \omega_{\pm}(t) \frac{\partial}{\partial \theta}\,.
  \label{eq:transformed_H}
\end{equation}

To simulate the dynamics under the Hamiltonian in Eq.~\eqref{eq:transformed_H}, we have found the simple split-propagator method~\cite{FeitJCP1982, KosloffJCP1988} to be effective. The results of such simulations, which use 1024 spatial grid points to represent the wave packets in the range $[0, \pi/4]$ and a time resolution of $\Delta t = \qty{1}{\microsecond}$, are shown in Fig.~\ref{fig:cn-sim-results} as the points labeled ``SP''. An excellent agreement with the inertial-frame results from the CN method is observed. We have also verified the precision of the split-propagator method by comparing it to a Chebychev propagation~\cite{Tal-EzerJCP1984}, which is exact to machine precision, but slower by about a factor of four. It is noted in Fig.~4 that the implementation with faster ramps, cf.\ panel (b), is still adiabatic. The faster ramp allows a longer loop time, accommodating $n=10$ cycles instead of just 2 within the same interferometer time $T$, and thus results in a higher sensitivity.

\subsection{Path integral method}%
\label{subsec:path}

The interferometric response closely follows a semiclassical model based on path-integral propagators. The propagator phase of a wave packet equals $\exp(iS(\textbf{x}_0(t))/\hbar) $, where $S(\textbf{x}_0(t))$ is the action of the classical trajectory, $\textbf{x}_0(t)$, followed by the centroid of the wave packet. Consequently, the phase difference $\Delta\Phi_S$ between our relevant pair of wave packets in spin states $\ket{-}$ and $\ket{+}$, arises from the difference of corresponding actions,
  \begin{equation}
    \Delta \phi_S = \frac{1}{\hbar}\int \left(\mathcal{L}(\textbf{x}_{-}, \dot{\textbf{x}}_{-}, t) - \mathcal{L}(\textbf{x}_{+}, \dot{\textbf{x}}_{+}, t)  \right) dt\,.
  \label{eq:generic-sc-phase}
\end{equation}
where $\textbf{x}_\pm$ are the paths followed by the centroids of the split wave-function components, and $ \mathcal{L}(\textbf{x}_{\pm}, \dot{\textbf{x}}_{\pm}, t)$ are the corresponding Lagrangians. In TAI, the predetermined lattice trajectories serve as the classical paths since the atomic wave functions remain tightly trapped at the minima of the relatively slowly-moving lattice potentials, and the wave functions possess zero degrees of freedom. That is, the $\textbf{x}_\pm$ are simply given by the locations of the selected sites of the optical lattices for $\ket{+}$ and $\ket{-}$, and forces of constraint cause no significant alterations. The Lagrangians for the states $\ket{+}$ and $\ket{-}$  differ in the presence of a non-zero background angular velocity $\Omega$ due to the different lattice angular speeds in the inertial frame. In the semi-classical path-integral picture, Eq.~(\ref{eq:generic-sc-phase}) leads to the well-known Sagnac phase,
\begin{equation}
  \Delta \Phi_S = \frac{4 m_{\text{Rb}} \Omega A}{\hbar}\,,
  \quad
  A = \frac{R^2}{2} \int_{0}^{T}\!\omega(t^\prime)\,dt^\prime\,.
  \label{eq:sagnac}
\end{equation}

The final recombined population in the state $\ket{-}$ on its respective potential,  Eq.~\eqref{eq:pop_minus} with $\Delta\Phi = \Delta\Phi_S$, is shown in Fig.~\ref{fig:cn-sim-results} as the solid curve. Figure~4~(a) shows the result for $\omega_0 = \qty{10}{\pps}$ and two cycles (the minimum number of cycles possible for $t_r = \qty{100}{ms}$ and $\tLoop >0$, for a total of $T=\qty{300}{ms}$). The parameters for Fig.~\ref{fig:cn-sim-results}~(b) match those for Fig.~\ref{fig:adiabatic_dynamics}. The close agreement of the semi-classical results with the full quantum simulations (CN and SP) in both Figs.~\ref{fig:cn-sim-results}~(a) and~(b) validates the principles of TAI in the adiabatic limit, in which unwanted spin couplings, wave-packet excitation and tunneling on the spin-dependent lattice potentials do not affect the interferometric phase of the TAI.%

\subsection{Rotation sensitivity}%
\label{subsec:sensitivity}

Assuming that a phase resolution of ${2\pi}/100$ can be experimentally achieved~\cite{duspayev_tractor_2021}, in
Fig.~\ref{fig:cn-sim-results}~(a) the rotation sensitivity can be inferred to be about \qty{5}{mrad/s}. This can be improved by increasing the lattice angular velocity for a given duration of the interferometric scheme. Fig.~\ref{fig:cn-sim-results}~(b) shows the response for $n=10$ cycles, achieved by increasing $\omega_0$ to $\qty{50}{\pps}$. In
Fig.~\ref{fig:cn-sim-results}~(b) the sensitivity is improved five-fold to roughly \qty{1}{mrad/s}.

The interferometer sensitivity can be enhanced further by increasing both the lattice radius $R$ and the angular velocity $\omega_\pm$ to an extent where centrifugal force and nonadiabatic excitation still remain negligible. In a lattice of depth $h\times \qty{4.4}{MHz}$ the orbital radius suffers $<0.1\%$ increase when rotated at $\omega_\pm \sim \qty{1000}{\pps}$. Nonadiabatic excitations are better explained in the co-rotating lattice frames given by Eq.~(\ref{eq:transform}). The first two terms of the Hamiltonian $\tilde{H}_\pm(t)$ in Eq.~\eqref{eq:transformed_H} describe stationary optical lattices, in which we initialize the wave functions in the respective ground states. As the lattices are accelerated, the last term, which is proportional to $\omega_\pm(t)\hat{L}_z$, may cause nonadiabatic transitions into excited vibrational states within the initially populated lattice wells. In shallow lattices, modified tunneling behavior may occur (Bloch oscillations and Wannier-Stark localization). In the following, we develop an estimate as to what rotation sensitivities may be possible under these constraints.

The departure from perfect adiabaticity can be quantitatively estimated in the momentum picture by exploiting the spatial periodicity of the Hamiltonian  $\tilde{H}_\pm(t)$. Following the well-known Bloch formalism, any eigenstate of $\tilde{H}_\pm(t)$ can be characterized by quasi-angular momentum $\ell$ and band index $n$ as $\ket{\psi^n_\ell} = e^{i\ell\theta}\ket{u^n_\ell} $ where $u^n_\ell(\theta+2\pi)=u^n_\ell(\theta)$. Then the effective Hamiltonian for $\ket{u^n_\ell}$ is given by
\begin{equation}
    \mathcal{H}_{\ell,\pm(t)} = \frac{(L_z+\hbar \ell)^2}{2I} + V_0 \cos\left(m\theta\right) + \omega_{\pm}(L_z+\hbar \ell)\,,
    \label{eq:k-space-transformed-H}
\end{equation}
where $\mathcal{H}_{\ell,\pm(t)} u_\ell(t) = E^n_\ell(t)u_\ell(t)$. Scaling the Hamiltonian by and effective recoil energy $E_R = \hbar^2 m_{Rb}^2/2I$ gives a dimensionless eigenvalue equation in terms of $\Theta = m\theta$,
\begin{widetext}
\begin{equation}
    \left[-\frac{d^2}{d\Theta^2} - \frac{2i\ell}{m}\frac{d}{d\Theta} + \left(\frac{\ell}{m}\right)^2 + \frac{V_0}{E_R}\cos(\Theta) - \frac{2I\omega_{\pm}}{m\hbar}\left(-i\frac{d}{d\Theta}+\frac{\ell}{m}\right)  \right]u^n_\ell(\omega_{\pm},\Theta) = \frac{E^n_\ell}{E_R} u^n_\ell(\omega_{\pm},\Theta)\,.
  \label{eq:scaled-transformed-H}
\end{equation}
\end{widetext}

Eq.~(\ref{eq:scaled-transformed-H}) offers an estimate of the relative magnitudes of different terms in the Hamiltonian in the co-rotating frames. First, we consider tunneling effects for a ground-state wave function $u^0_0(0,\theta)$ trapped in a static lattice (\ie, $ \ell=\omega_{\pm}=0 $). In this simple case, tunneling-induced wave-function delocalization is suppressed when $V_0 \gg E_R$. When a lattice of such potential depth is rotated at a constant angular velocity $\omega_{\pm}$, the final term on the left-hand side in Eq.~(\ref{eq:scaled-transformed-H}) mixes the ground state of the stationary lattice with excited states from higher bands. This mixing can be minimized if the lattice depth $2V_0$ is much larger than the scale of the lattice-rotation-induced perturbation  $m \hbar \omega_\pm$.

For the pinwheel lattice under consideration, the effective recoil energy $E_R$ is $ \sim h~\times$~\qty{250}{Hz} and the scale of the lattice-rotation-induced perturbation at $\omega_\pm \simeq \qty{500}{\pps}$ is $\sim h\times \qty{80}{kHz} $.  Therefore, a lattice depth of $h \times$ \qty{4.4}{MHz} can adequately suppress delocalization of the ground-state wave function and support a maximum angular velocity up to $\omega_{\pm} \sim \qty{500}{\pps}$ with minimal nonadiabatic excitations. Pinwheel lattices of radius \qty{2.5}{mm} rotated at \qty{1000}{\pps} can potentially improve the sensitivity of Fig.~\ref{fig:cn-sim-results}~(b) by six orders of magnitude to \qty{1}{nrad/s} for an operation time of \qty{1}{s}.

The signal-to-noise ratio can be enhanced by loading a larger number of atoms into the lattices. This can be achieved, for instance, by creating pinwheel lattices with more sites (\ie, larger $m$) and stacking several pinwheel lattices axially on a linear array of $z$-lattice sites. The axial stacking is limited to a range $|z| \ll z_R $, where the locations of the radial minima stay similar enough to avoid excessive inhomogeneous broadening of the TAI phase $\Delta \phi$. For example, in Fig.~\ref{fig:pinwheel-lattice-potential}~(a), the position of the radial minima changes by less than 0.01\% over $z=\pm$\qty{1.5}{\micrometer}. Therefore, for a stack of 10 pinwheel lattices separated by the lattice period \qty{266}{nm} one can expect the average TAI fringe contrast to remain large for $\Delta \phi$ up to several $100 \times 2 \pi$. In experimental realizations, LG beams with larger beam waists and Rayleigh ranges would allow a higher degree of axial stacking to improve signal-to-noise.

\section{Nonadiabatic effects in lattice spin-up and -down}%
\label{sec:nonadiabatic}

\begin{figure}[tb]
 \centering
  \includegraphics{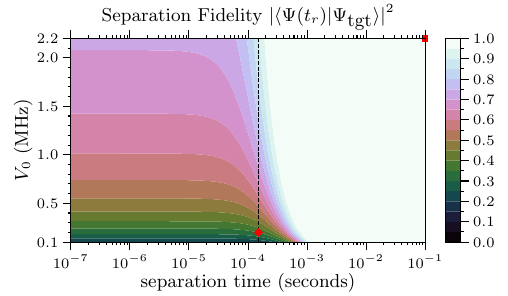}
  \caption{%
    \label{fig:fidelity_map}
    Fidelity of the initial splitting operation for varying separation time $t_r$ and $V_0$ of the trapping potential. The separation fidelity is the overlap of the state $\ket{\Psi(t_r)}$ resulting from the evolution under Eq.~\eqref{eq:drive_up} with the ground state of the moving potential at $t_r$.
  }
\end{figure}

In order to further optimize the gyroscope sensitivity and to increase the dynamic range in rotation sensing, the time $t_r$ in Eq.~(\ref{eq:drive}) during which the lattices are accelerated should be reduced. Additionally,
the ability of the interferometer to operate with
shallower lattices, which will accommodate laser power constraints and minimize signal loss due to photon scattering, has to be explored. When entering the nonadiabatic regime, the split wave function in each spin-dependent potential deviates from the ground state $\Psi_0(\theta)$ in the lattice rest frames. We have studied the nonadiabatic effects numerically by simulating the time evolution under the Hamiltonian in Eq.~\eqref{eq:transformed_H}. In Fig.~\ref{fig:fidelity_map},
we show the fidelity under the smoothly-varying ramp function $\omega(t)$ in Eq.~\eqref{eq:drive_up}, which drives the ground state $\Psi_0(\theta, t=0)$ in the initially selected optical lattice well into a state $\Psi(\theta, t_r)$. The fidelity is given by the magnitude-square of the overlap between
$\Psi(\theta, t_r)$ and the desired target state,
$\Psi_{\tgt}(\theta, t_r)$, which is the ground state of the potential rotating at the terminal constant speed $\pm \omega_0 = \qty{50}{\pps}$. The point marked by the red square in the top-right corner corresponds to the fully adiabatic time evolution shown in Fig.~\ref{fig:adiabatic_dynamics}.

\begin{figure*}[tb]
 \centering
  \includegraphics{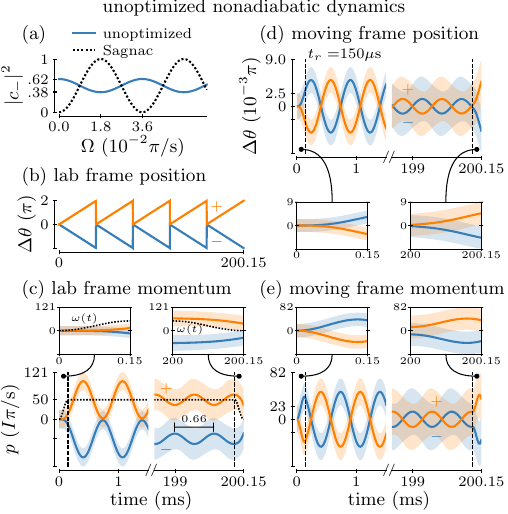}
  \includegraphics{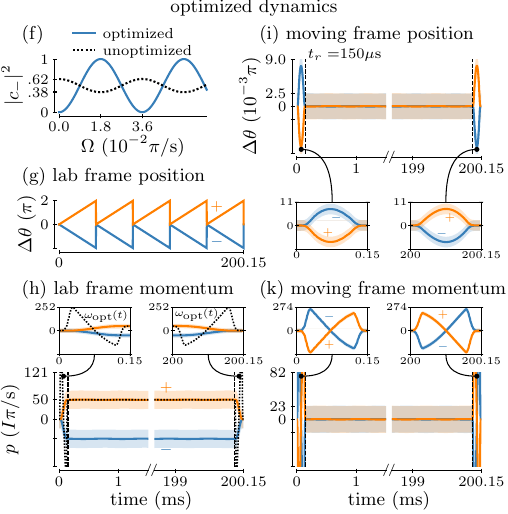}
  \caption{%
    \label{fig:opt_dynamics}
    Dynamics of the TAI interferometer for a nonadiabatic separation time $t_r = \qty{150}{\microsecond}$ and a shallow potential with $V_0 = \qty{0.2}{MHz}$, as marked by the red diamond in Fig.~\ref{fig:fidelity_map}. All other parameters are as in Fig.~\ref{fig:adiabatic_dynamics}. Panels (b--f) show the results under the analytic drive function $\omega(t)$ given by Eq.~\eqref{eq:drive}, and panels (g--k) show the results for an optimized field $\omega_{\opt}(t)$ (see text for details). Panels (a, f) show the interferometric response of the un-optimized and the optimized drive functions to a constant background rotation $\Omega$. For comparison, the semiclassical Sagnac curve from Fig.~\ref{fig:cn-sim-results}~(b) is included in panel~(a).
  }
\end{figure*}

We observe a transition from adiabatic to nonadiabatic evolution for a separation time between \qty{1}{ms} and \qty{100}{\microsecond}, depending on the depth of the potential. It is thus confirmed that the lattice acceleration conditions in Fig.~\ref{fig:adiabatic_dynamics} are deep in the adiabatic regime, allowing several orders of magnitude increase in acceleration before nonadiabatic effects actually become substantial. To gain a better understanding of the separation failure mode for small $t_r$ and $V_0$ and of the effects of nonadiabaticity on the overall interferometric scheme, we show in Fig.~\ref{fig:opt_dynamics}~(a)-(e) the dynamics for $t_r = \qty{150}{\microsecond}$ and $V_0 = \qty{0.2}{MHz}$, marked with the red diamond in Fig.~\ref{fig:fidelity_map}. Looking first at the initial separation phase, see left insets in Fig.~\ref{fig:opt_dynamics}~(c)-(e), we can see that the wave packet is not readily accelerated to terminal speed by the accelerating optical lattice. The lab-frame momentum, shown in panel~(c), shows very little initial acceleration of the wave packet. Unlike in the adiabatic case in Fig.~\ref{fig:adiabatic_dynamics}~(b), where $\Avg{p}_{+}$ readily reaches $\omega_0 = \qty{50}{\pps}$ at $t=t_r$, in Fig.~6~(c) it
does not even come close. In the moving frame (lattice rest frame), shown in Fig.~6~(d) and~(e), both position and momentum are far from zero, which also contrasts against the adiabatic case in Fig.~\ref{fig:adiabatic_dynamics}~(c) and~(d). In fact, initially the signs of momentum and position in the moving frame are opposite to that of the acceleration: as the lattice is accelerated to the left (counter-clockwise), the wave packet in the moving frame is displaced to the right (clockwise).

To explain the observations in the previous paragraph, we first note that during the short separation time of \qty{150}{\microsecond} the actual displacement is very small: note the scale factor of $10^{-3}$ on the y-axis in Fig.~\ref{fig:opt_dynamics}~(d).
In the subsequent \qty{199.85}{ms}, when the optical lattices loop at constant counter-rotating speeds $\pm \omega_0$, the atoms eventually respond to the force that the trapping potential imparts on them.
As a consequence, in the lattice rest frames the wave packets oscillate around zero, as seen in the panels~(d) and~(e), and around momentum $\pm\omega_0$ in the lab frame, as seen in panel~(c).
Visual inspection of Fig.~\ref{fig:opt_dynamics}~(c) reveals that the oscillation period is close to that of the harmonic $\omega_H$ in Eq.~\eqref{eq:omega_H}, which is \qty{660}{\microsecond} for $V_0 = \qty{0.2}{MHz}$.
However, the atoms oscillate in the anharmonic regions of the cosine potential in Eq.~\eqref{eq:hamiltonian}, for this value of $V_0$.
As a result, we also find a breathing of the oscillation, \ie, the oscillation amplitude diminishes while the width of the wave packet increases, see the first and last \qty{1.5}{ms} in panels~(c) and~(e).
The breathing is absent when the atoms remain confined to the near-harmonic sections of the cosine function, \eg, for potentials with larger values of $V_0$.
The physical picture that summarizes and underlies these observations is that for $t_r \rightarrow 0$ and $V_0$ sufficiently large, the lattices instantaneously speed up to $\omega_0$ underneath the atoms.
If the corresponding kinetic energy in the lattice frame is less than $2V_0$, the atoms subsequently undergo a sloshing oscillation in the lattice frame.
At longer times, the oscillation exhibits collapse and quantum revival phenomena caused by the anharmonicity of the potential.

In Figs.~6~(a-e), the ramp-down of the pinwheel lattice from $\omega_0$ to a position at rest, in the lab frame, behaves fundamentally the same as the initial ramp-up: the wave packet does not slow down with the rapidly decelerating potential. Instead, the location of the oscillating wave packet at the time instant when the deceleration hits determines the wave packet's state within the lattice well after the lattice slowdown is complete. The final state may range from less excited to more highly excited than before the deceleration. In Fig.~\ref{fig:opt_dynamics}~(d) and~(e) the latter is the case.

Overall, the lack of fidelity seen in Fig.~\ref{fig:fidelity_map}, and the resulting oscillatory dynamics have a detrimental effect on the contrast of the full TAI interferometric scheme. First, as shown in Fig.~\ref{fig:opt_dynamics}~(b), the interferometer fails to close perfectly. Second, prior to the final recombination $\pi/2$ pulse the wave packets no longer match the ground state $\Psi_0(\theta)$, neither in position nor in momentum, and nor in width, as seen in panels~(d) and~(e). Thus, the magnitude of the overlap $\eta$ in Eq.~\eqref{eq:eta_overlap} typically is much less than 1, and the contrast of the resulting populations $|c_{\pm}|^2$ in Eq.~\eqref{eq:pop_nonadiabatic} is correspondingly diminished. This result is shown in Fig.~\ref{fig:opt_dynamics}~(a). For the given parameters, the achieved contrast is only 24\%. This falls well short of the contrast of the path-integral Sagnac curve, shown as the black dotted line, which matches exactly Fig.~\ref{fig:cn-sim-results}~(b). Because for small $t_r$ the overall process approximates the physics of two impulsive kicks applied to a wave-packet in a well, there also is an erratic dependence of the AI contrast on fine details. The phase of the wave-packet sloshing motion at the time instant of the second kick largely determines visibility. The simplified two-pulse picture becomes more accurate at shorter $t_r$; the picture essentially applies in the left third of Fig.~5. An additional factor that plays a role is that at small $V_0$ and short $t_r$, the effective two-pulse wave-packet drive may excite the wave packet partially into the continuum, causing further contrast loss. In the next section, we will attempt to correct these unwanted behaviors using methods of optimal control.

\section{Optimal Control}%
\label{sec:optimalcontrol}

Having observed the detrimental effect of a separation time $t_r$ that is too short, we consider the use of optimal control to improve the fidelity in Fig.~\ref{fig:fidelity_map} for moderate values of $t_r$. Specifically, we seek to find an $\omega(t)$ that is an alternative to the analytical shape in Eq.~\eqref{eq:drive_up} such that $\Psi_{\pm}(\theta, t)$ reaches $\Psi_{\pm, \tgt}(\theta, t=t_r)$, that is, the ground state in the lattice rest frames. The optimized $\omega(t)$ must maintain the boundary conditions $\omega(0) = 0$ and $\omega(t_r) = \omega_0$. To this end, we parametrize
\begin{equation}
  \omega_{\opt}(t) = \omega(t) + S(t)\delta\omega(t)\,,
\end{equation}
where $\omega(t)$ is the original shape given by Eq.~\eqref{eq:drive_up}, $\delta\omega(t)$ is a correction to be optimized, and $S(t) \in [0, 1]$ is a fixed shape with $S(0) = S(t_r) = 0$ to enforce the boundary conditions. Here, we use a shape that smoothly switches on and off with a Blackman shape during the first and last 20\% of the time window. The initialization for $\delta\omega(t)$ is $\delta\omega(t) = 0$.

An optimized correction $\delta\omega(t)$ can be obtained using any of the standard gradient-based quantum control methods, including  GRAPE~\cite{KhanejaJMR2005} or Krotov's method~\cite{Tannor1992}. Here, we have used the Krotov.jl package~\cite{Krotov.jl} within the QuantumControl Julia framework~\cite{QuantumControl.jl}. Within 300 iterations, using a square-modulus functional~\cite{PalaoPRA2003}, we can bring the separation error from 0.648, see Fig.~\ref{fig:fidelity_map}, to $1.26 \times 10^{-5}$.
The resulting optimized $\omega_{\opt}(t)$ is shown in the left inset of Fig.~\ref{fig:opt_dynamics}~(h), with the full resulting dynamics for the entire interferometric scheme in panels~(g)-(k). The optimized control function for the ramp-down is the time inverse of the ramp-up one, see the right inset of panel~(h).

We observe a ``throw and catch'' behavior. The field ramps up rapidly to a relatively high (but still achievable) speed of \qty{252}{\pps}, but then slows down and temporarily switches direction, before returning to the target speed of \qty{50}{\pps}. The lab frame momentum does not follow this rapid motion, but smoothly accelerates from 0 to \qty{50}{\pps}, as can be seen in the inset of Fig.~\ref{fig:opt_dynamics}~(h), and very much mimics the adiabatic dynamics in Fig.~\ref{fig:adiabatic_dynamics}~(b). Likewise, the lattice-frame position, seen in panel~(i), initially lags behind the accelerating lattice potential, but then smoothly catches up to the equilibrium position within the lattice frame. The subsequent dynamics while the optical lattices loop at constant speed $\omega_0$ are near-identical with the adiabatic case in Fig.~\ref{fig:adiabatic_dynamics}: both position and momentum are zero in the moving frame, see panels~(i) and~(k),  and follow the position and momentum of the trapping potential in the lab frame, see  panels~(g) and~(h).  The ramp-down inverts the dynamics during the ramp-up, leaving the wave function in a state that is very close to the ground state of the lattice potential at rest. This results in near-ideal interferometric response following Eq.~\eqref{eq:pop_adiabatic}, as shown in Fig.~\ref{fig:opt_dynamics}~(f). This implies that the interferometric path in panel~(g) is now perfectly closed, in contrast to the open path in panel~(b).

In principle, a ``throw and catch'' optimal control solution can be found for even shorter $t_r$. However, the shorter $t_r$, the larger the amplitude that the control function $\omega_{opt}(t)$ will need to reach during the ramp-up and ramp-down phases. Hence, the maximum experimentally achievable  angular control velocity will determine how far one may push to the left in Fig.~\ref{fig:fidelity_map}.

A more general approach to accelerate the ramp-up and ramp-down phases is to exploit the dependency of the boundary between adiabatic and nonadiabatic behavior on $V_0$ (see Fig.~\ref{fig:fidelity_map}). For instance, starting from the point marked by the red diamond in Fig.~\ref{fig:fidelity_map}, one may \emph{temporarily} increase the depth of the potential, move up along the dashed line in Fig.~\ref{fig:fidelity_map}, in combination with tuning $\omega(t)$. However, an eigenstate of a
shallower lattice well will not be an eigenstate of a deeper one, resulting in a breathing motion of the wave packet after the compression. To counter this, one would have to add another layer of ``throw and catch'' to suppress the breathing, or introduce additional control over the shape of the potential.

The control functions we have obtained here are already quite simple and can be readily implemented. As an alternative or an augmentation to the numerically optimized controls, one may in the future explore \emph{analytic} control schemes under the umbrella of ``shortcuts to adiabaticity''~\cite{Guery-OdelinRMP2019}.

\section{Conclusion}%
\label{sec:conclusion}

In summary, we have presented the design of a rotation sensor based on the principles of tractor atom interferometry~\cite{duspayev_tractor_2021, raithel2022principles}. An experimental setup can be realized using readily available instrumentation. The parameters for the pinwheel lattice, which is at the heart of the envisioned devices, can be obtained following the considerations discussed in Sec.~\ref{sec:design}. In the adiabatic limit, quantum-dynamics simulations of the sensor's single- and multi-loop operation agree well with semi-classical path-integral predictions, in which one simply enters the known tractor trajectories into the applicable Lagrangian. Our proof-of-principle simulations allow rotation sensitives of about 1~mrad/s. We have provided a discussion and concrete examples that illustrate the utility of quantum control to realize fast beam splitters and ramps to prepare coherently-split wave-function components that counter-rotate at high rotation speeds, allowing higher sensitivity. Estimates that extrapolate pinwheel area and rotation speed to reasonable limits predict sensitivities approaching 1~nrad/s, at a 1-second measurement time. Future investigations may further explore the benefits of quantum entanglement~\cite{salvi2018, anders2021, CarrascoPRA2022, greve2022} for increasing the sensitivity-bandwidth product, reducing sensor size etc. Moreover, optimal control theory techniques, as utilized here to reduce the splitting time and to alleviate the influence of nonadiabatic effects and decoherence caused by photon scattering, may present a viable pathway to improve the performance of matter-wave interferometers, including future experiments at the International Space Station~\cite{Frye2021, alonso2022}.

\section*{ACKNOWLEDGMENTS}
We thank Ansh Shah for useful discussions and initial computational work. The work at the University of Michigan was supported by the Army Research Office and DEVCOM Army Research Laboratory under Cooperative Agreement Number W911NF-2220155, and by the NSF Grant No. PHY-2110049. AD acknowledges support from the Rackham Predoctoral Fellowship at the University of Michigan. MHG and SCC acknowledge support by the DEVCOM Army Research Laboratory under Cooperative Agreement Number W911NF-16-2-0147 and W911NF-21-2-0037, respectively. VSM is grateful for support by a Laboratory University Collaboration Initiative (LUCI) grant from OUSD.

\bibliography{references.bib}

\newpage
\appendix
\onecolumngrid

\section{Generalized Crout Reduction}
\label{sec:pbc-Crout}
The time evolution of the wave function under Schrödinger's equation is given by
\begin{equation}
  \frac{\partial \psi(x, t)}{\partial t}= \frac{i \hbar}{2 m} \frac{\partial^{2} \psi(x, t)}{\partial x^{2}} - \frac{i V(x)}{\hbar} \psi(x, t)\,.
\end{equation}
Let us consider this in a 1-D system confined to $\left[-x_l, x_l\right)$ with the initial conditions $\psi(x,0) = \psi^0(x)$, $\psi(x_l,0) = \psi(-x_l, 0) = 0$ and periodic boundary condition so that $\psi(x+2x_l)=\psi(x)$. By discretizing the derivatives, the right-hand-side is approximated up to $\mathcal{O}(\dx^2)$ as
\begin{equation}\label{H_def}
  \frac{i \hbar}{2 m \dx^{2}}\left(\psi_{j+1}^{(k)}+\psi_{j-1}^{(k)}-2 \psi_{j}^{(k)}\right)- \frac{iV_j}{\hbar} \psi_{j}^{(k)}
  \equiv \sum_{m=0}^{N} \frac{i H_{j, m} \psi_{m}^{(k)}}{\hbar}\,,
\end{equation}
where $H$ is the position representation of the Hamiltonian. The Crank-Nicolson method discretizes the time domain by taking the average of forward and backward differences to approximate the time derivative $\frac{\partial \psi}{\partial t}$; and relates $\psi^{k+1}\equiv \psi(t=(k+1)\delta t)$ to $\psi^k$ as
\begin{equation}\label{full_CN}
  \left(1-\frac{i H \Delta t}{2\hbar}\right) \psi^{k+1} = \left(1+\frac{i H \Delta t}{2\hbar}\right) \psi^{k}\,.
\end{equation}



The generalized Crout reduction algorithm for the PBC Hamiltonian is:
\begin{outline}[enumerate]
  \1 Initialize $ \psi^{k+1}, l, u, z, \eta, \xi $ as zero arrays of size $ N $
  \1 With  $\lambda = \frac{i\hbar \Delta t}{2m\Delta x^2}$
     \2 Set $l_0 = 1+ \lambda + \frac{i V_0 \Delta t}{2\hbar} $, $ \eta_0 = -\frac{\lambda}{2} $, $ \xi_0 =  -\frac{\lambda}{2l_0} $
    \2 For $ i $ in $ \{1,\dots,N-2\} $:
    \begin{equation*}
        \begin{split}
            u_i &= -\frac{\lambda}{2l_i} \\
            l_i &= 1+ \lambda + \frac{i V_i \Delta t}{2\hbar} + \frac{\lambda}{2}u_i
        \end{split}
    \end{equation*}

    \2  For $ i $ in $ \{1,\dots,N-2\} $ :
      $$ \xi_i = \frac{\lambda^{i+1}}{2^{i+1}\prod_{j=0}^{i}l_j}, \quad \eta_i = \frac{\lambda^{i+1}}{2^{i+1}\prod_{j=0}^{i-1}l_j} $$
    \2 Set the last entries of $ l $ and $ u $:
      \begin{equation*}
          \begin{split}
            u_{N-1} &= -\frac{\lambda}{2l_{N-2}} + \xi_{N-2}\\
            l_{N-1} &= 1+ \lambda + \frac{i V_{N-1} \Delta t}
            {2\hbar}\\ &- \sum_{j=0}^{N-3} \eta_j\xi_j - \left(\eta_{N-2} - \frac{\lambda}{2} \right)u_{N-1}
          \end{split}
      \end{equation*}
      \vspace{1pt}
  \1 Set $ z_0 = \left(1 - \lambda - \frac{i V_0 \Delta t}{2\hbar}\right)\psi^k_0 +\frac{\lambda}{2l_0}\left(\psi^k_{N-1}+\psi^k_1 \right)  $
    \2 For $ i $ in $ \{1,\dots,N-2\} $:
    \begin{equation*}
    \begin{split}
        z_i = &\left(1 - \lambda - \frac{i V_i \Delta t}{2\hbar}\right)\psi^k_i +\frac{\lambda}{2l_i}\left(\psi^k_{i+1}+\psi^k_{i-1}+z_{i-1} \right)
    \end{split}
    \end{equation*}
    \2 Set last entry:
      \begin{equation*}
                \begin{split}
                     z_{N-1} =& \left(1 - \lambda - \frac{i V_{N-1} \Delta t}{2\hbar}\right)\psi^k_0\\ +&\frac{\lambda}{2l_{N-1}}\left(\psi^k_{N-1}+\psi^k_1+z_{N-2} - \sum_{j=0}^{N-2}z_j\eta_j \right)
                \end{split}
      \end{equation*}

  \1 Back-substitute to obtain $ \psi^{k+1} $:
    \2 Set $ \psi^{k+1}_{N-1} = z_{N-1} $
    \2 For $ i $ in $ \{N-2,\dots,1\} $:
    $$ \psi^{k+1}_i  = z_i + \frac{\lambda}{2l_i}\psi^{k+1}_{i+1} - \psi^{k+1}_{N-1}\xi_i $$
\end{outline}

\end{document}